\documentclass[aps,prl,reprint,twocolumn,amsmath,amssymb,groupedaddress]{revtex4-1}
\usepackage{graphicx}
\usepackage[hidelinks,breaklinks=true]{hyperref}

\begin{document}

\title{Stochastic Dynamics of Nanoparticle and Virus Uptake}

\author{Felix Frey}
\author{Falko Ziebert}
\author{Ulrich S. Schwarz}

\affiliation{Institute for Theoretical Physics, Heidelberg University, Philosophenweg 19, 69120 Heidelberg, Germany 
and BioQuant, Heidelberg University, Im Neuenheimer Feld 267, 69120 Heidelberg, Germany}

\date{\today}

\begin{abstract}
The cellular uptake of nanoparticles or viruses requires that the gain of adhesion energy
overcomes the cost of plasma membrane bending. It is well known that this leads to a minimal
particle size for uptake. Using a simple deterministic theory for this process, we first show
that, for the same radius and volume, cylindrical particles should be taken up faster than 
spherical particles, both for normal and parallel orientations. 
We then address stochastic effects, which are expected
to be relevant due to small system size, and show that, now, spherical particles
can have a faster uptake, because the mean first passage time profits from the multiplicative noise
induced by the spherical geometry. We conclude that stochastic effects are strongly 
geometry dependent and may favor spherical shapes during adhesion-driven particle uptake.
\end{abstract}

\maketitle

Biological cells constantly communicate with their environment by relaying signals
and material across their plasma membranes. In particular, cells routinely take up 
small particles with sizes in the order of $10-300\,{\rm nm}$. This process
is usually driven by receptor-ligand interactions \cite{alberts}, such that the adhesion
energy can overcome the energy required for membrane bending. 
It is also exploited by viruses that have to enter host cells for replication \cite{virus_entry}.
Due to the physical nature of this process, cells also take up 
artificial nanoparticles of various shapes \cite{nano_up},
whose uptake may be either intentional or undesired, as in
drug delivery  \cite{drugdelivery} or in case of microplastics \cite{Moos}, respectively.

\begin{figure}
\includegraphics[ width=.44 \textwidth]{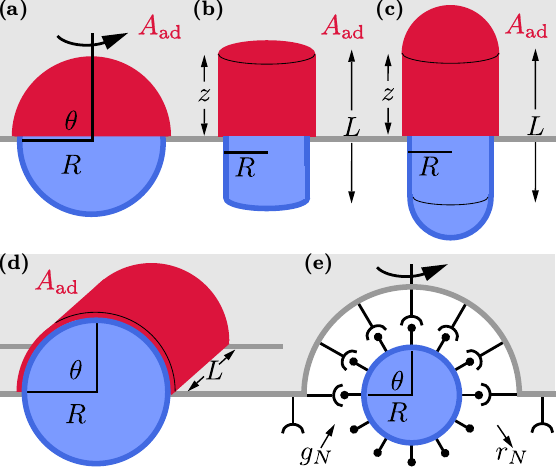}
\caption{Uptake of a particle of (a) spherical, (b) normal cylindrical (\textit{rocket mode}),
(c) spherocylindrical and
(d) parallel cylindrical shape (\textit{submarine mode}). In a deterministic model, 
the virus is homogeneously covered with ligands, 
adhering to the cell membrane along the adhesive area $A_\mathrm{ad}$.
(e) Stochastic uptake of a virus, for which the surface presents discrete ligands. 
The virus particle binds (unbinds) with forward rate $g_N$ (backward rate $r_N$) 
to receptors on the cell membrane.}
\label{fig:Figure1}
\end{figure}

Viruses come in many different shapes \cite{hulo2010viralzone}, 
but the most paradigmatic and most frequently occuring ones are
spherical and cylindrical \cite{crick1956structure,roos2010physical}.
Therefore here we focus our discussion on spheres, cylinders and a combination of both,
as shown in Fig.~\ref{fig:Figure1}.
It is generally believed that the spherical shape is optimal because
it maximizes the volume to surface ratio and confers high mechanical stability \cite{BruinsmaPNAS2004}.
However, it is less clear which shapes are optimal in regard to
interactions with the membrane \cite{nano_up}. Here we show that stochasticity might play
an important role in this context.

Due to their small size, viruses are typically covered with only 
few tens of ligands for cell surface receptors \cite{kumberger2016,sun2006}
and thus stochastic effects are expected to be relevant. For example, the icosahedral and
medium-sized ($60-100\,{\rm nm}$) members of the family of reoviruses have only $12$ primary JAM-A-$\sigma1$ (junction adhesion molecule) 
ligands on their surfaces \cite{barton2001}. Although stochastic effects have
been argued to be essential for the stability of small adhesion clusters \cite{ErdmannSchwarzPRL,ROBIN2018},
particle uptake is usually studied analytically only with deterministic approaches.
Stochastic effects are automatically included in computer simulations with thermostats \cite{vacha2011receptor,huang2013role}, but
such approaches do not allow for a systematic study of the role of noise.
Here we analytically calculate the first passage times for particle uptake of various shapes and show that
they strongly depend on geometry. Our main result is that in contrast 
to deterministic dynamics, stochastic dynamics tends to favor spherical shapes for uptake.

\begin{figure*}
\includegraphics[ width=1.\textwidth]{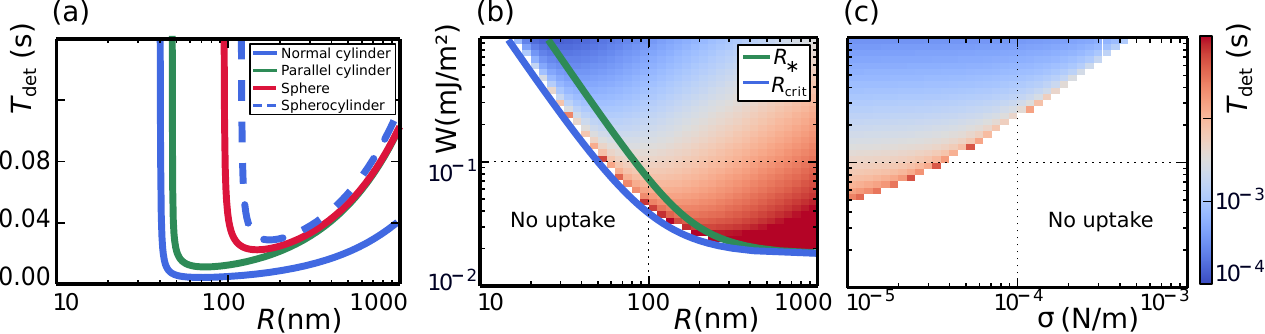}
\caption{Deterministic uptake dynamics. (a) Uptake times for sphere (red), normal cylinder (blue), spherocylinder (dashed blue)
and parallel cylinder (green) as function of radius $R$ at equal particle volume.  
(b), (c) Uptake times for a sphere as a function of adhesion energy $W$ and radius $R$
or membrane tension $\sigma$, respectively. 
In (b) the critical (optimal) radius for spherical uptake is shown in blue (green). 
Parameter values (if not varied) are 
$R=90\,{\rm nm}$, $W=4.4\cdot10^{-2}\,{\rm mJ/m}^2$ and 
$\sigma=0.9\cdot 10^{-5}\,{\rm N/m}$.
}
\label{fig:Figure2}
\end{figure*}

Deterministic approaches usually start by balancing the contributions from adhesion
and bending. Because adhesion energy scales with the radius squared,
while bending energy is independent of radius, a minimal size exists for particle uptake \cite{LipoDoeb}. 
The most investigated aspect of receptor-mediated uptake is the role of particle 
shape \cite{huang2013role,dasgupta2014,Bahrami}.
More detailed approaches include the variational problem for 
finding minimal energy shapes \cite{DesernoPRE04,agudo2015}, 
the role of receptor diffusion towards the nanoparticle \cite{gao2005,decuzzi2007role},
particle deformability \cite{yi2011cellular,zeng2017contact}, the physics of the scission step \cite{mcdargh2016constriction} and the role of the
cytoskeleton \cite{sun2006}. 
In order to develop a stochastic description, here we start with
a minimal deterministic model, which we then extended to the stochastic case.

We assume that ligands to the cellular adhesive receptors
are homogeneously distributed on the viral surface.
The total free energy can be written as \cite{helfrich1973} 
\begin{equation}
E_{\mathrm{total}}=-\int_{A_\mathrm{ad}} W \mathrm{d} A
+ \int_{A_\mathrm{mem}}  2\kappa H^2  \mathrm{d}A + \sigma \Delta A
\label{eq:free_energy_general}
\end{equation}
where $W$ is the adhesion energy per area, $\kappa$ is the bending rigidity, 
$H$ is the mean curvature, $\sigma$ is the membrane tension and $\Delta A$ 
the excess area due to the deformation (compared to the flat case). 
Typical parameter values are 
$W=10^{-1}\,{\rm mJ/m}^2$, $\kappa=25\,{\rm k_{\mathrm{B}}T}$ and $\sigma=10^{-4}\,{\rm N/m}$
\cite{foret2014,kumar2016}. For example, a reovirus has $12$ major and $60$ 
minor ligands \cite{barton2001,veesler2014}.
With a binding energy of around $15\,{\rm k_{\mathrm{B}}T}$ each, 
one estimates $W=4.4\cdot10^{-2}\,{\rm mJ/m}^2$ for a homogeneous adhesion model. 

\begin{figure*}
\includegraphics[ width=1.\textwidth]{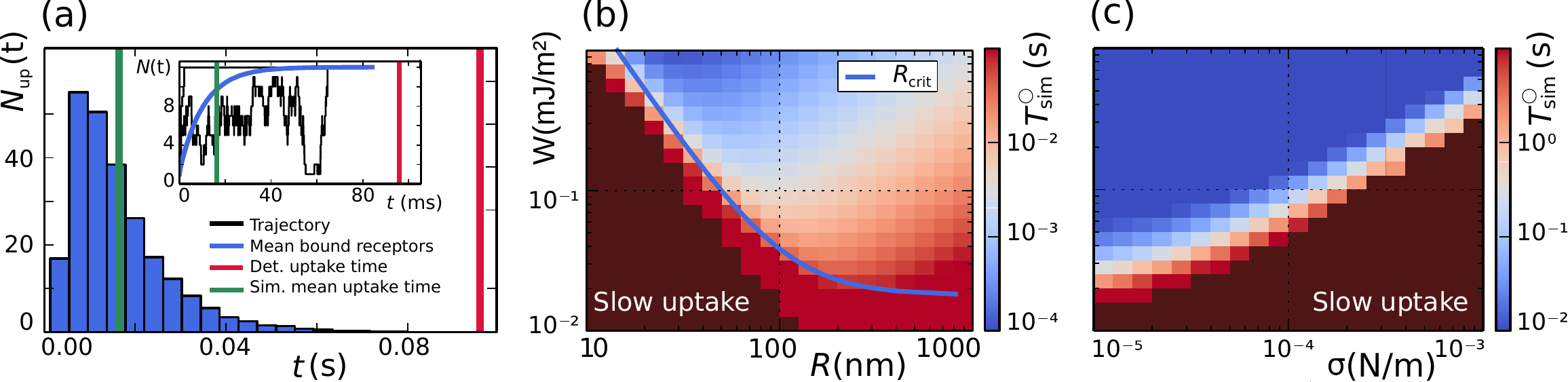}
\caption{Stochastic uptake dynamics for a spherical particle (reovirus with $N_{\mathrm{max}}=12$).
    (a) Histogram of uptake times with obtained mean 
    $ T_{\mathrm{sim}}^{\circ}\approx 16\,{\rm ms}$ (green line;
    standard deviation is $\approx11\,{\rm ms}$) as compared to the deterministic
    value $T_{\mathrm{det}}^{\circ} \approx 96\,{\rm ms}$ (red line). 
    (Inset) Two example trajectories (black) of the number of bound receptors as a function of time 
    and the mean number $\langle N^{\mathrm{sim}} \rangle$ obtained from 
    simulating the master equation (blue).
    (b), (c) Mean uptake time $T_{\mathrm{sim}}^{\circ}$ 
    as a function of $W$ and either $R$ or $\sigma$. 
    In the dark colored region, uptake may still occur beyond the used simulation time.
    In (b) the blue line is the critical radius $R_{\mathrm{crit}}$ of the deterministic model.
    Parameter values as in Fig.~\ref{fig:Figure2}.
}
\label{fig:Figure3}
\end{figure*}

The two membrane parameters define a typical length scale 
$\lambda = \sqrt{\kappa / \sigma} \approx 32\,{\rm nm}$. 
As shown schematically in Fig.~\ref{fig:Figure1}, we consider 
a spherical or cylindrical particle of radius $R$ adhering to the membrane.
Then the bending energy in Eq.~(\ref{eq:free_energy_general}) has contributions 
both from the adhering ($A_\mathrm{ad}$) and free part 
($A_\mathrm{mem}-A_\mathrm{ad}$) of the membrane. 
It has been previously shown \cite{foret2014,kumar2016,dasgupta2014,sadeghi2018}
that the contributions from the free part can be neglected in two special regimes.
In the loose regime ($R\ll \lambda$), the free membrane can adopt the
shape of a minimal surface and thus its bending contribution vanishes.
In the tense regime ($R\gg \lambda$), the free part is essentially flat
and the contribution from the adhered membrane is much larger
than the one from the free membrane. To arrive at an analytical model, here we neglect the contributions from the free membrane
also for the intermediate case.

Eq.~(\ref{eq:free_energy_general}) can now easily be analyzed for the geometries sketched in  
Fig.~\ref{fig:Figure1}, namely (a) for a sphere ($\circ$), (b) for a cylinder oriented normally to the membrane ($\bot$), 
(c) for a spherocylinder ($\cap$) and (d) for a cylinder oriented parallel to 
the membrane ($\parallel$). 
Although it has been shown in coarse-grained molecular dynamics simulations that
spherocylinders might undergo a dynamical sequence of first lying down and then 
standing up \cite{huang2013role}, the two cylindrical modes considered here 
are the paradigmatic configurations encountered during wrapping \cite{dasgupta2014}.
In case of the normal cylinder, the top and bottom surfaces are neglected as they do not contribute to the uptake force.
To keep our calculations as transparent as possible, we neglect them also for the parallel cylinder.
In both cases, we neglect the bending energies at the kinked edges.

The uptake forces $F_{\mathrm{up}}$ are calculated by taking the variation 
of the energy with respect to opening angle $\theta$ or height $z$, respectively,
and are balanced by the friction force
they experience, $F_{\mathrm{up}}=F_{\mathrm{friction}}$ \cite{agudo2015}.
For a spherical particle, the membrane covered area is a spherical cap 
of radius $R$ and opening angle $\theta$, 
i.e.~$A_{\mathrm{ad}}^{\circ}=2\pi R^2  (1-\cos\theta)$, cf.~Fig.~\ref{fig:Figure1}(a). 
The friction force is 
$F_{\mathrm{friction}}^{\circ}  = \eta 2 \pi R \sin(\theta)  R \dot{\theta}$, i.e.\  
an effective membrane microviscosity of order $\eta=1\ {\rm Pa \; s}$ times the change
of the membrane-covered particle surface. 
The resulting equation of motion reads 
\begin{equation}
\dot{\theta}=
\nu_{\mathrm{up}}-\nu_{\sigma} (1-\cos\theta)
\label{eq:det_ODE}
\end{equation}
with $\nu_{\mathrm{w}} = W/(R \eta)$, $\nu_{\mathrm{\kappa}} = 2 \kappa/(R^3 \eta)$,
$\nu_{\sigma}=\sigma/(R\eta)$ and $\nu_\mathrm{up}=  \nu_{\mathrm{w}}-\nu_{\mathrm{\kappa}}$.
The uptake time can be calculated as \cite{suppl}
\begin{equation}
T_{\mathrm{det}}^{\circ}
\approx
\frac{\pi}{ \nu_{\mathrm{up}}  \sqrt{1-\frac{2 \nu_{\mathrm{\sigma}}}{\nu_{\mathrm{up}} }}}\ .
\end{equation}
Note that it diverges for $\nu_{\mathrm{up}}= 2\nu_{\mathrm{\sigma}}$, 
defining a critical radius $R_{\mathrm{crit}}$,
below which uptake is not possible. In the limit of vanishing $\sigma$, we recover the classical result
$R_{\mathrm{crit}} = \sqrt{2 \kappa / W}\approx 44\,{\rm nm}$ \cite{LipoDoeb}.
We note that our theory predicts normal uptake forces of around ten pN, which
agrees well with results from recent atomic force microscopy experiments 
\cite{alsteens2017nanomechanical,pan2017process}.

Analogous calculations can be performed for the three cases with cylinders at equal volumes \cite{suppl}.
Fig.~\ref{fig:Figure2}(a) compares the resulting uptake times.
For the normal and parallel cylinders, we take the same radius as for the sphere and adjust
the aspect ratio. For the spherocylinder, we take the same aspect ratio as for the cylinders
and adjust the radius. 
All four geometries show similar behaviors: a critical radius $R_{\mathrm{crit}}$ exists, below which uptake is not possible.
The parallel cylinder has half the critical radius of the sphere because
its mean curvature is half as large at equal radius.
Moreover an optimal radius $R_{*}$ exists, at which the uptake time is 
minimal \cite{suppl}. Interestingly, the critical and optimal values are very close to each other,
and the cylindrical particles are taken up faster than the spherical ones.
The spherocylinder is the slowest case, because
at equal volumes, the aspect ratio is modest and the spherical part dominates.
Fig.~\ref{fig:Figure2}(b) and (c) display the uptake time 
for a spherical particle as a function of $W$ and $R$ or $\sigma$, respectively,
showing that a smaller adhesion energy has to be compensated by
either larger radius or smaller membrane tension. Importantly, in the deterministic
case certain parameter values do not lead to any uptake.

We now turn to the stochastic variant of our uptake model (cf.~Fig.~\ref{fig:Figure1}(e)). 
For the sphere, we map the membrane covered area onto the number of bound receptors $N$ \cite{suppl},
assuming axial symmetry. Using Eq.~(\ref{eq:det_ODE}), we find a 
dynamical equation for $N$ through $\dot{N}=(\mathrm{d} N/\mathrm{d} \theta) \dot{\theta}$.
From this discrete equation a one-step master equation \cite{vankampen1992} can then be deduced, 
with the forward rate $g_N=\nu_{\mathrm{w}} N_{\mathcal{E}}$ and the backward rate
$r_N=\nu_{\mathrm{\kappa}} N_{\mathcal{E}}+2 \nu_{\mathrm{\sigma}} N_{\mathcal{E}}(N-1)/(N_{\mathrm{max}}-1)$,
where $N_{\mathcal{E}  }(N) =\sqrt{(N-1)[(N_{\mathrm{max}}-1)-(N-1)]}$ corresponds to the
number of receptors at the advancing edge.

We first solved the master equation numerically using the Gillespie algorithm \cite{gillespie1977} 
and averaging over $10^4$ trajectories.
Fig.~\ref{fig:Figure3}(a) shows the resulting distribution of uptake times
and the results for the number of bound receptors as a function of time (inset).
Clearly, the mean uptake time is substantially smaller than 
the uptake time from the deterministic approach. 
Fig.~\ref{fig:Figure3}(b) and (c) display the simulated mean uptake times
as a function of $W$ and $R$ or $\sigma$, respectively. 
Comparing to the deterministic approach, cf.~Fig.~\ref{fig:Figure2}(b) and (c),
we see that now uptake is possible for any parameter value, although
it can be rather long in regions where the deterministic
model does not allow for uptake. However, the parameter region with uptake in 
experimentally accessible uptake times is strongly increased and 
now also extends below the blue line
indicating the critical radius $R_{\mathrm{crit}}$ of the deterministic model. 
This expansion of the parameter regime
that allows the process to complete is also known from the stochastic
dynamics of small adhesion clusters \cite{ErdmannSchwarzPRL,ROBIN2018}. 

\begin{figure*}
\includegraphics[ width=1.\textwidth]{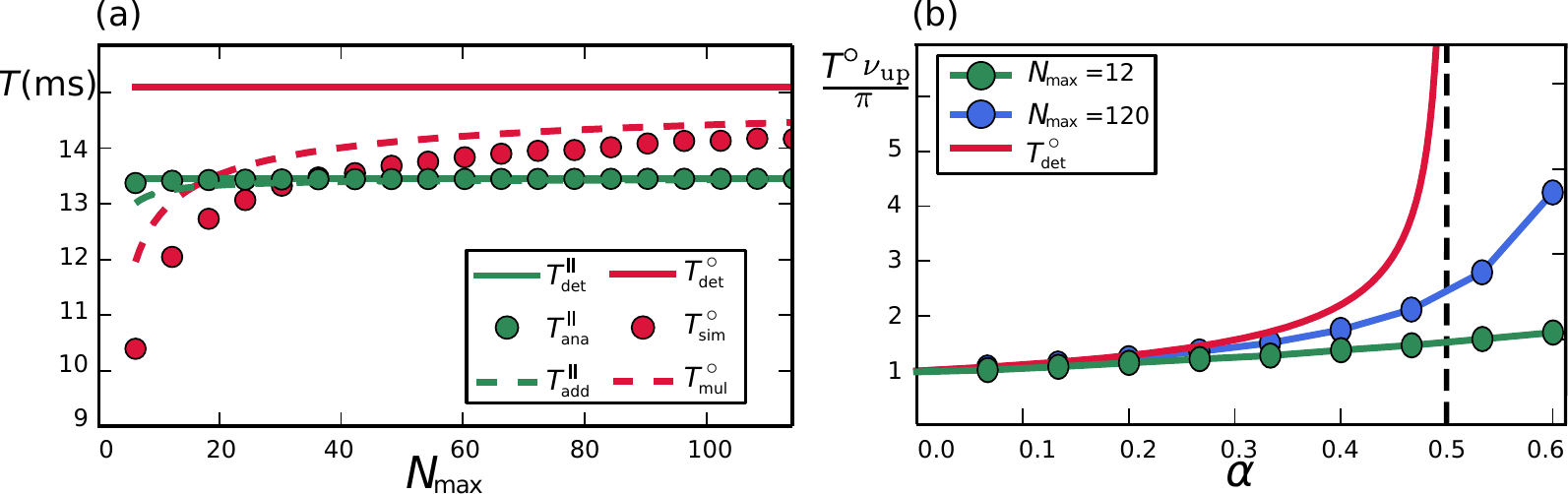}
\caption{(a) Geometry-dependent mean uptake times for spheres (red) and 
    parallel cylinders (green) as a function of the maximum number of receptors.
    Shown are the analytical results for the deterministic case (solid) and 
    for multiplicative (additive) noise corresponding to the spherical (cylindrical) geometry (dashed). 
    The result from the simulations of the master equation is shown for sphere (cylinder) as symbols. 
    (b) The case with membrane tension can be treated with computer simulations.
    Shown are the mean uptake times for a sphere as a function of the dimensionless parameter
    $\alpha=\nu_{\mathrm{\sigma}}/\nu_{\mathrm{up}}$ (by varying $\sigma$)
    for two different numbers of receptors and the deterministic case 
    (diverging at $\alpha=1/2$).
    }
\label{fig:Figure4}
\end{figure*}

We next discuss the interplay between shape and stochastic dynamics in analytical detail.
For simplicity we set the membrane tension to zero in the following ($\sigma =0$). 
We approximate the master equation by a Fokker-Planck equation (FPE)
using a Kramers-Moyal expansion \cite{vankampen1992}.
The equivalent stochastic differential equation
can be transformed to angle space
using It\^{o}'s Lemma \cite{gardiner1985}
\begin{equation}
\dot{\theta}=\nu_\mathrm{up} - D \frac{\cos\theta}{\sin^2\theta} + \sqrt{\frac{2D}{\sin\theta}}\,\xi(t)
\label{eq:SDE_continuous}
\end{equation}
where $D=(\nu_{\mathrm{w}}+\nu_{\mathrm{\kappa}})/(N_{\mathrm{max}}-1)$. The noise $\xi(t)$ is assumed 
to be Gaussian with
$\left < \xi(t)\right > = 0 $ and $\left < \xi(t) \xi(t^\prime )\right > = \delta (t-t^\prime)$.
From Eq.~(\ref{eq:SDE_continuous}) one can directly read the drift term (or
the corresponding potential) of the corresponding FPE in angle space 
and its diffusion coefficient $D$ \cite{suppl}. Because 
for the spherical case this diffusion constant depends on
the state variable $\theta$, here we deal with multiplicative noise.

In marked contrast, for the two cylindrical cases one obtains additive noise.
For example, for the parallel cylinder we find
$\dot{\theta}^{\parallel}=\nu_\mathrm{up}^{\parallel} + \sqrt{2D^{\parallel}}\, \xi(t)$
where $\nu_{\mathrm{up}}^{\parallel}=W/(\eta R)-\kappa/(2 R^3 \eta)$ and 
$D^{\parallel}=(\nu_{\mathrm{up}}^{\parallel}+\kappa/(R^3 \eta))
\pi/(2 (N_{\mathrm{max}}-1))$,
where the latter does not depend on $\theta$ 
because the length of the the moving contact line is constant \cite{suppl}.
The different quality of the noise suggests that the uptake dynamics 
change in a fundamental manner in the different geometries.

The mean uptake times can be obtained analytically studying the mean first passage time problem
using the backwards FPE \cite{gardiner1985,redner2001}  
with reflecting (adsorbing) boundary condition at $\theta=0$ ($\theta=\pi$). 
Neglecting the angle dependent drift term (as it is large only for the first and last step)
but keeping the multiplicative noise, the mean uptake time evaluates to \cite{suppl}
\begin{equation}
\label{result_mult}
T_{\mathrm{mult}}^{\circ}=
T_{\mathrm{det}}^{\circ}
\left [ 1-\mathrm{e}^ {-\frac{\nu_{\mathrm{up}}}{D}}  I_0 \left(\frac{\nu_{\mathrm{up}}}{D} \right) \right]
 < T_{\mathrm{det}}^{\circ}
\end{equation}
where $I_0$ is the modified Bessel function of the first kind.
In the limit of small fluctuations  compared to the uptake frequency, 
we recover the deterministic limit. In the opposite limit of large fluctuations,
the uptake time approaches $T_{\mathrm{mult}}^{\circ}\approx \pi / D$.

For the parallel cylinder the noise is additive and the mean uptake time is readily obtained as \cite{suppl}
\begin{align}
\label{result_add}
 T_{\mathrm{add}}^{\parallel} 
 = T_{\mathrm{det}}^{\parallel}
 -\frac{ D^{\parallel}}
 { {\nu_{\mathrm{up}}^{\mathrm{\parallel\,2}}}}
 \left[1-\exp \left(-\frac{\pi  \nu_{\mathrm{up}}^{\parallel}}{ D^{\parallel}}\right)\right] 
 < T_{\mathrm{det}}^{\parallel}
\end{align}
where $T_{\mathrm{det}}^{\parallel} =\pi / \nu_{\mathrm{up}}^{\parallel}  $.
In the limit of small fluctuations
one again recovers the deterministic uptake time, while for large fluctuations one finds
$T_{\mathrm{add}}^{\parallel}\approx \pi^2/(2 D^{\parallel})$. 
Hence, for both geometries the mean uptake time is always smaller than the deterministic one.
In general, fluctuations in small systems combined with a reflecting boundary 
should always decrease the mean first passage time, since the stochastic process 
profits from the presence of the boundary, while the deterministic process 
does not.

We now consider a particle with $R=180\,{\rm nm}$, i.e.~in the region of Fig.~\ref{fig:Figure2}(a) 
where the deterministic uptake times of sphere and parallel cylinder are similar.
Fig.~\ref{fig:Figure4}(a) shows the mean uptake times for different geometries
at equal volume as a function of the number of receptors. 
We note that for the parallel cylinder it is also possible to 
compute the uptake time directly from the master equation $T_\mathrm{ana}^{\parallel}$ \cite{gardiner1985}.
The agreement between simulations (symbols) and analytical results (lines) is very good for cylinders and
rather good for spheres. For small number of receptors, i.e.~strong
fluctuations, the uptake of a sphere is faster than the one of a cylinder.
We conclude that uptake of spherical particles dynamically benefits 
from the noise. In fact, using in Eq.~(\ref{result_add}) $\nu_{\mathrm{up}}$ and $D$ 
instead of $\nu_{\mathrm{up}}^{\parallel}$ and $D^{\parallel}$,
we find that $T_{\mathrm{mult}}^{\circ}<T_{\mathrm{add}}^{\circ}<T_{\mathrm{det}}^{\circ}$ always holds.

While membrane tension could not be treated analytically,
it can be included in the simulations, and 
we get the same results, i.e.~the uptake times are reduced by increasing stochasticity.
Fig.~\ref{fig:Figure4}(b)
shows the mean uptake times as a function of the dimensionless parameter
$\alpha=\nu_{\mathrm{\sigma}}/\nu_{\mathrm{up}}$
for different 
$N_{\mathrm{max}}$
and $R=90\,{\rm nm}$. Although stochasticity is most important for small numbers of 
receptors, nevertheless,
even for substantial numbers on the order of one hundred receptors,
the stochastic effects survive.

In conclusion, we found that the uptake of spherical particles 
profits from the presence of noise. Our results suggest
yet another advantange for viruses to be spherical. Similar effects arising from the
interplay of stochastic dynamics and geometry might also
exist for other biologically relevant first passage problems,  
e.g.~phagocytosis \cite{Tollis}, the closure of transient pores 
on spherical vesicles \cite{sandre1999}
or the fusion of tissue over circular holes \cite{nier2015}.

\newpage

\begin{acknowledgments}
F.F. acknowledges support by the Heidelberg Graduate School for 
Fundamental Physics (HGSFP).
The authors thank Steeve Boulant, Ada Cavalcanti-Adam and Tina Wiegand for 
helpful discussions on reovirus, and we acknowledge the Collaborative Research 
Centre 1129 for support.
U.S.S. acknowledges support as a member of the 
Interdisciplinary Center for Scientific Computing
and the cluster of excellence CellNetworks. 
\end{acknowledgments}


%

\end{document}